\documentclass[12pt]{article}
\usepackage{graphicx}
\usepackage{amsmath}

\newcommand\pubnumber{}
\newcommand\pubdate{\today}

\def\wayne{Department of Physics and Astronomy\\
Wayne State University, Detroit, MI 48201, USA}
\def\mctp{Michigan Center for Theoretical Physics, Department of Physics\\
University of Michigan, Ann Arbor, MI 48109, USA}
\def\fermilab{Theoretical Physics Department\\
Fermilab, P.O. Box 500, Batavia, IL 60510, USA}
\def\warsaw{Institute of Theoretical Physics\\
University of Warsaw, Pasteura 5, 02-093 Warsaw, Poland}
\def\support{\footnote{Speaker.}}

\def\Title#1{\begin{center} {\Large #1 } \end{center}}
\def\Author#1{\begin{center}{ \sc #1} \end{center}}
\def\Address#1{\begin{center}{ \it #1} \end{center}}

\newcommand\pubblock{\rightline{\begin{tabular}{l} \pubnumber\\
         \pubdate  \end{tabular}}}
\newenvironment{Abstract}{\begin{quotation}  }{\end{quotation}}
\newenvironment{Presented}{\begin{quotation} \begin{center} 
             PRESENTED AT\end{center}\bigskip 
      \begin{center}\begin{large}}{\end{large}\end{center} \end{quotation}}
\def\Acknowledgements{\bigskip  \bigskip \begin{center} \begin{large}
             \bf ACKNOWLEDGEMENTS \end{large}\end{center}}
\newcommand{\diff}{\mathrm{d}}
\newcommand{\Ord}[1]{\mathcal{O}\left(#1\right)}
\newcommand{\msbar}{\overline{\text{MS}}}
\def\mom{{\cal M}}

\newcommand{\alsmz}{\alpha_s(m_Z)}
\def\mom{{\cal M}}
\newcommand{\Obs}[2]{\widehat{O}_{#1}^{\text{#2}}}
\newcommand{\Oin}[1]{\{\widehat{O}_{#1}^{\text{in}}\}}
\newcommand{\inp}{\{\mathcal{I}_k\}}
\newcommand{\GamQ}{\Gamma_{H\to Q\bar Q}}
\newcommand{\Gamc}{\Gamma_{H\to c\bar c}}
\newcommand{\Gamb}{\Gamma_{H\to b\bar b}}
\newcommand{\mua}{\mu_\alpha}
\newcommand{\mum}{\mu_m}
\newcommand{\mumin}{\mu_{\rm min}}
\newcommand{\mumax}{\mu_{\rm max}}

\def\beq{\begin{equation}}
\def\eeq#1{\label{#1}\end{equation}}
\def\eeqn{\end{equation}}

\def\beqa{\begin{eqnarray}}
\def\eeqa#1{\label{#1}\end{eqnarray}}
\def\eeqan{\end{eqnarray}}

\def\leqn#1{(\ref{#1})}

\let\bar=\overbar

\def\Dslash{\not{\hbox{\kern-4pt $D$}}}
\def\dslash{\not{\hbox{\kern-2pt $\del$}}}

\def\ee{e^+e^-}

\def\msb{{\bar{\ssstyle M \kern -1pt S}}}

\begin{document}
\begin{titlepage}
\pubblock

\vfill
\Title{Resolving charm and bottom quark masses in precision Higgs boson analyses}
\vfill
\Author{Alexey~A.~Petrov$^{a,b,c}$%\footnote{Supported by the U.S.\ Department of Energy, Fermilab's Intensity Frontier Fellowship and URA Visiting Scholar Award.}%\footnote{Supported in part by DoE under contract DESC0007983, Fermilab's Intensity Frontier Fellowship and URA Visiting Scholar Award \#14-S-23.}
, 
Stefan~Pokorski$^d$%\footnote{Supported by the National Science Center in Poland.}%\footnote{Supported by the National Science Center in Poland under the research grants DEC-2012/05/B/ST2/02597 and DEC-2012/04/A/ST2/00099.}
, 
James~D.~Wells$^b$%\footnote{Supported by the U.S.\ Department of Energy.}%\footnote{Supported in part by DoE under grant DE-SC0011719.}
, 
Zhengkang~Zhang$^b$\support%\footnote{Supported by the U.S.\ Department of Energy.}%%\footnote{Speaker. Supported in part by DoE under grant DE-SC0011719.}
}
%\Author{Alexey A. Petrov\support}
% put in address(es) defined above
\Address{$^{(a)}$\wayne\\$^{(b)}$\mctp\\$^{(c)}$\fermilab\\$^{(d)}$\warsaw}
\vfill
\begin{Abstract}
Masses of the charm and bottom quarks are important inputs to precision calculations of Higgs boson observables, such as its partial widths and branching fractions. They constitute a major source of theory uncertainties that needs to be better understood and reduced in light of future high-precision measurements. Conventionally, Higgs boson observables are calculated in terms of $m_c$ and $m_b$, whose values are obtained by averaging over many extractions from low-energy data. This approach may ultimately be unsatisfactory, since $m_c$ and $m_b$ as single numbers hide various sources of uncertainties involved in their extractions some of which call for more careful estimations, and also hide correlations with additional inputs such as $\alpha_s$. Aiming at a more detailed understanding of the uncertainties from $m_c$ and $m_b$ in precision Higgs boson analyses, we present a calculation of Higgs boson observables in terms of low-energy observables, which reveals concrete sources of uncertainties that challenge sub-percent-level calculations of Higgs boson partial widths.
\end{Abstract}
\vfill
\begin{Presented}
The 7th International Workshop on Charm Physics (CHARM 2015)\\
Detroit, MI, 18-22 May, 2015
\end{Presented}
\vfill
\end{titlepage}
\def\thefootnote{\fnsymbol{footnote}}
\setcounter{footnote}{0}
%

%%%%%%%%%%%%%%%%%%%%%%%%%%%%%%%%%%
\section{Introduction}

Precision studies of the Higgs boson are crucial for understanding the origin of electroweak symmetry breaking and searching for new physics beyond the Standard Model (SM). Many well-motivated new physics scenarios predict percent-level deviations from the SM for the Higgs couplings~\cite{Gupta:2012mi}. Accordingly, it is hoped that data at a later stage of the LHC or future facilities will enable (sub)percent-level determinations of the Higgs boson branching fractions and the important partial widths $\GamQ$ ($Q=c,b$)~\cite{Asner:2013psa,Peskin:2013xra,Fan:2014vta,Ruan:2014xxa}. However, this does not necessarily imply that percent-level new physics effects can be probed. In fact, the power of precision Higgs analyses is limited by theory uncertainties, which are dominated by the parametric uncertainties from the input parameters, especially the charm and bottom quark masses~\cite{Denner:2011mq,Almeida:2013jfa,Lepage:2014fla}.\footnote{The perturbative uncertainty for $\GamQ$ is well below 1\%, thanks to the N$^4$LO calculation~\cite{Baikov:2005rw}.} In terms of the scale-invariant masses in the $\msbar$ scheme [i.e.~solutions to $m_Q(\mu)=\mu$], the uncertainty propagation, according to Ref.~\cite{Almeida:2013jfa},
\beq
\frac{\Delta\Gamc}{\Gamc} \simeq \frac{\Delta m_c(m_c)}{10~\text{MeV}} \times 2.1\%,\quad \frac{\Delta\Gamb}{\Gamb} \simeq \frac{\Delta m_b(m_b)}{10~\text{MeV}} \times 0.56\%,
\eeqn
indicates that this parametric uncertainty is at the level of a few percent at present, if the input quark masses are taken from the PDG particle listings~\cite{Agashe:2014kda},
\beq
m_c(m_c) = 1.275(25)\text{ GeV},\quad m_b(m_b) = 4.18(3)\text{ GeV}.
\eeq{mqpdg}
Note also that $\Delta\Gamb$ propagates into the calculations of all branching fractions since $b\bar b$ is the dominant decay channel of the SM Higgs boson. 
In light of the projected experimental precision on the Higgs observables, it is thus highly desirable to have a detailed understanding of this dominant theory uncertainty, and hence the improvements needed. In particular, instead of treating $m_c$ and $m_b$ as single numbers to be read from the PDG, we will make an initial attempt to incorporate the extraction of $m_c$, $m_b$ from low-energy experiments into precision Higgs analyses. As a result, the vague notion of ``uncertainties from quark masses'' will be decomposed into concrete sources, some of which represent a serious challenge that calls for further investigation of the low-energy observables. Further details of this work can be found in Ref.~\cite{Petrov:2015jea}.

%%%%%%%%%%%%%%%%%%%%%%%%%%%%%%%%%%
\section{A global approach to precision Higgs analyses}

A conventional approach to precision Higgs analyses is to regard $m_c$ and $m_b$ as input ``observables'', namely to put them on the same footings as $m_Z$ and $m_H$. As such they are assigned ``experimental'' central values and error bars, as in Eq.~\leqn{mqpdg}. However, unlike $m_Z$ and $m_H$, quark masses are not well-defined observables due to confinement. Instead, they are just parameters in the SM Lagrangian. Their values are extracted from the true observables whose theory predictions depend on them. In fact, the PDG quark masses quoted in Eq.~\leqn{mqpdg} are obtained by averaging over many results in the literature, which are extracted from a variety of observables. Much information is hidden by this averaging procedure, making the conventional approach of using the numbers in Eq.~\leqn{mqpdg} unsatisfactory in several aspects. First, the averaging unrealistically assumes no correlations among the various quark masses extractions, some of which use similar data and/or methods. Second, there are correlations between $m_Q$ and $\alpha_s$ because $\alpha_s$ enters the quark masses extractions as an input, but such correlations are not retained in Eq.~\leqn{mqpdg}. $m_Q$ and $\alpha_s$ are then treated as independent inputs when the Higgs observables are calculated, which is strictly speaking incorrect. Furthermore, the meaning of the error bars in Eq.~\leqn{mqpdg} is obscure. They contain not only various experimental and theoretical uncertainties associated with many different observables, but also a self-described inflation of uncertainties by the PDG (see the ``Quark masses'' review in Ref.~\cite{Agashe:2014kda}). The latter is introduced to account for possibly underestimated uncertainties in some of the reported quark masses in the literature. The problem of uncertainty underestimation was first noticed in Ref.~\cite{Dehnadi:2011gc}, and its impact on precision Higgs analyses will be discussed below.

These unsatisfactory aspects of the conventional approach can in principle be eliminated if we include only true {\it observables} $\{\Obs{i}{}\}$ in the $\chi^2$ fit, where the $\chi^2$ function
\beq
\chi^2 = \sum_{ij} \biggl[\Obs{i}{th}(\inp) - \Obs{i}{expt}\biggr] V^{-1}_{ij} \biggl[\Obs{j}{th}(\inp) - \Obs{j}{expt}\biggr],
\eeqn
with appropriately determined uncertainties and correlations contained in the covariance matrix $V$, is minimized with respect to the inputs of the calculation $\inp$. $m_c$ and $m_b$ are in the set $\inp$, but not in the set $\{\Obs{i}{}\}$.

For such a fit to be useful, the fit observables $\{\Obs{i}{}\}$ should include those contributing to the extraction of $m_c$ and $m_b$. Interestingly, while our starting point is precision Higgs analyses, the observables dominating the PDG average of $m_Q$ are associated with much lower energy scales $E\sim m_Q \ll m_H$. They include, for example, low~\cite{Dehnadi:2011gc,Kuhn:2001dm,Kuhn:2007vp,Chetyrkin:2009fv,Dehnadi:2015fra} and high~\cite{Signer:2008da,Hoang:2012us,Penin:2014zaa,Beneke:2014pta} moments of $\ee\to Q\bar Q$ inclusive cross section, defined by
\beq
\mom_n^Q \equiv \int\frac{\diff s}{s^{n+1}}R_Q(s),\quad\text{where}\,\, R_Q\equiv\frac{\sigma(e^+e^-\to Q\bar QX)}{\sigma(e^+e^-\to \mu^+\mu^-)},
\eeq{momdef}
variants of these moments~\cite{Bodenstein:2011ma,Bodenstein:2011fv}, and moments of lepton energy and hadron mass distributions of semileptonic $B$ decay~\cite{Gambino:2013rza,Buchmuller:2005zv,Bauer:2004ve}.

As a result, we are led to the following picture:
\beq
\begin{Bmatrix}
\Obs{1}{low}(m_c, m_b, \alpha_s, \dots)\\[0.5ex]
\Obs{2}{low}(m_c, m_b, \alpha_s, \dots)\\[0.5ex]
\Obs{3}{low}(m_c, m_b, \alpha_s, \dots)\\[0.5ex]
\vdots
\end{Bmatrix}
\Leftarrow
\begin{Bmatrix}
\text{\underline{Inputs}}\\
m_c\\
m_b\\
\alpha_s\\
\vdots
\end{Bmatrix}
\Rightarrow
\begin{Bmatrix}
\Obs{1}{Higgs}(m_c, m_b, \alpha_s, \dots)\\[0.5ex]
\Obs{2}{Higgs}(m_c, m_b, \alpha_s, \dots)\\[0.5ex]
\Obs{3}{Higgs}(m_c, m_b, \alpha_s, \dots)\\[0.5ex]
\vdots
\end{Bmatrix}.
\eeqn
The low-energy observables $\{\Obs{i}{low}\}$ play an important role in precision Higgs analyses, because they are sensitive to the same set of inputs as the Higgs observables $\{\Obs{i}{Higgs}\}$. The role of low-energy observables is not obvious in the conventional approach, where a large amount of information from $\{\Obs{i}{low}\}$ has been highly processed into just two numbers $m_c$ and $m_b$. As we strive for higher-precision calculations, this information should be resolved, and the low-energy observables should be more directly engaged. In this way it is in principle straightforward to include correlations among the observables, and retain the correct $\alpha_s$ dependence in each observable. We propose this global approach as a long-term goal for the precision Higgs analysis program, which will become more relevant as experimental precisions on $\{\Obs{i}{Higgs}\}$ improve over time, for both rigorous tests of the SM and fits to SM extensions. The observables set can also be enlarged to include precision electroweak observables (e.g.\ $Z$-pole observables, $m_W$, and LEP2 data) to make the global approach even more powerful.

%%%%%%%%%%%%%%%%%%%%%%%%%%%%%%%%%%
\section{Anatomy of theory uncertainties in $\Gamc$, $\Gamb$}

To assess the theory uncertainties in calculating $\{\Obs{i}{Higgs}\}$ without performing a full-fledged $\chi^2$ fit, in particular the uncertainties from our imprecise knowledge of $m_c$ and $m_b$, it is helpful to pick out two low-energy observables from the set $\{\Obs{i}{low}\}$, and eliminate $m_c$ and $m_b$ in favor of them in the functions $\Obs{i}{Higgs}(m_c, m_b, \alpha_s, \dots)$. We will choose $\mom_1^c$ and $\mom_2^b$, defined in Eq.~\leqn{momdef}, as the two low-energy observables, motivated by the simplicity of their calculations and the small quoted uncertainties for $m_c$ and $m_b$ extracted from them. These moments can be calculated by the method of relativistic QCD sum rules~\cite{Novikov:1977dq} (see e.g.\ Refs.~\cite{Shifman:1998rb,Colangelo:2000dp} for reviews)\footnote{We note in passing that the sum rules approach has been recently recast by the lattice QCD community~\cite{McNeile:2010ji,Chakraborty:2014aca,Colquhoun:2014ica}. See Ref.~\cite{Lepage:2014fla} for its possible impact on future precision Higgs analyses.}, which relate them to the vector current correlators:
\beq
\mom_n^Q = \frac{12\pi^2}{n!}\left(\frac{\diff}{\diff q^2}\right)^n\Pi_Q(q^2)\biggr|_{q^2=0},
\eeqn
where
\beq
(q^2g_{\mu\nu}-q_\mu q_\nu)\Pi_Q(q^2) = -i\int\diff^4x\,e^{iq\cdot x}\langle0|Tj_\mu(x)j_\nu^\dagger(0)|0\rangle.
\eeqn
$j_\mu$ is the electromagnetic current of the quark $Q(=c,b)$. $\Pi_Q$ can be calculated via an operator product expansion:
\beq
\mom_n^Q = \frac{\bigl(Q_Q/(2/3)\bigr)^2}{\bigl(2m_Q(\mum)\bigr)^{2n}} \sum_{i,a,b} C_{n,i}^{(a,b)}(n_f) \biggl(\frac{\alpha_s(\mua)}{\pi}\biggr)^i \ln^a\frac{m_Q(\mum)^2}{\mum^2} \ln^b\frac{m_Q(\mum)^2}{\mua^2} + \mom_n^{Q,\text{np}},
\eeq{sr}
where $Q_Q$ is the electric charge of the quark $Q$, and $C_{n,i}^{(a,b)}$ are functions of the number of active quark flavors $n_f$ (4 for $\mom_n^c$ and 5 for $\mom_n^b$). The two terms come from perturbation theory and nonperturbative condensates, respectively. Low moments (small $n$) are preferred so that the perturbative part, which has been calculated up to $\Ord{\alpha_s^3}$~\cite{Maier:2009fz}, dominates. The nonperturbative piece $\mom_n^{Q,\text{np}}$ is dominated by the gluon condensate contribution, which has been calculated to next-to-leading order~\cite{Broadhurst:1994qj}, and is nonnegligible only for the charm quark.

The moments $\mom_n^Q$ are functions of $m_Q$ and $\alpha_s$, which are renormalized at $\mum$ and $\mua$, respectively, in the $\msbar$ scheme. Physical observables like $\mom_n^Q$ should not depend on the renormalization scales $\mum$, $\mua$. But when they are calculated to finite order in perturbation theory there is residual scale dependence, which is conventionally used to estimate the effects of unknown higher-order terms. It used to be a common practice to set $\mum=\mua$, but it is argued in Ref.~\cite{Dehnadi:2011gc} that the perturbative uncertainty is generally underestimated in this way. Keeping $\mum$ and $\mua$ separate, we can invert Eq.~\leqn{sr} to extract $m_Q$, as has been done in Refs.~\cite{Dehnadi:2011gc,Dehnadi:2015fra},
\beqa
m_c(m_c) &=& m_c(m_c) \Bigl[\alsmz,\mom_1^c,\mum^c,\mua^c,\mom_1^{c,\text{np}}\Bigr],\label{mcextmu}\\
m_b(m_b) &=& m_b(m_b) \Bigl[\alsmz,\mom_2^b,\mum^b,\mua^b\Bigr].\label{mbextmu}
\eeqan
Here we have assumed the input $\alpha_s$ is at the scale $m_Z$, from which $\alpha_s(\mua)$ can be derived by the renormalization group (RG) equations~\cite{Chetyrkin:2000yt}. Also, RG equations allow us to convert the extracted $m_Q(\mum)$ to $m_Q(m_Q)$. It is clear that the quark masses depend not only on the low-energy observables $\mom_n^Q$, but also on $\alsmz$ and the renormalization scales in the calculation of $\mom_n^Q$. All this dependence is retained in Eqs.~\leqn{mcextmu} and~\leqn{mbextmu}, and is eventually propagated into the calculated Higgs observables in $\{\Obs{i}{Higgs}\}$. We will focus on the partial widths $\Gamc$, $\Gamb$ in the following. Neglecting correlations between $m_c$ and $\mom_1^{c,\text{np}}$ (which is justified given the large uncertainties in $\mom_1^{c,\text{np}}$, see Ref.~\cite{Petrov:2015jea}), we have
\beqa
\Gamma_{H\to c\bar c} &=& \Gamma_{H\to c\bar c} \Bigl[\Oin{k},m_c(m_c),\mu_H^c\Bigr]\nonumber\\
 &=& \Gamma_{H\to c\bar c} \Bigl[\Oin{k},\mom_1^c,\mum^c,\mua^c,\mu_H^c,\mom_1^{c,\text{np}}\Bigr],\label{Gamcmu}\\
\Gamma_{H\to b\bar b} &=& \Gamma_{H\to b\bar b} \Bigl[\Oin{k},m_b(m_b),\mu_H^b\Bigr]\nonumber\\
 &=& \Gamma_{H\to b\bar b} \Bigl[\Oin{k},\mom_2^b,\mum^b,\mua^b,\mu_H^b\Bigr]\label{Gambmu},
\eeqan
where $\Oin{k} \equiv \{ m_Z,\; G_F,\; \alpha(m_Z),\; m_t,\; \alsmz,\; m_H \}$ is the input observables set familiar in precision electroweak analyses. $\mu_H^Q\sim\Ord{m_H}$ collectively denotes the renormalization scales chosen in the calculations of $\Gamc$ and $\Gamb$, and should not be confused with $\mu_m^Q,\mu_\alpha^Q\sim\Ord{2m_Q}$. Note that the $\alsmz$ dependence has changed in the second equalities in Eqs.~\leqn{Gamcmu} and~\leqn{Gambmu} to account for the correlations between $\alpha_s$ and $m_Q$ reflected in Eqs.~\leqn{mcextmu} and~\leqn{mbextmu}.

Eqs.~\leqn{Gamcmu} and~\leqn{Gambmu} show the decomposition of the theory uncertainties in $\Gamc$, $\Gamb$. In particular, what people usually refer to as ``uncertainties from $m_c$ and $m_b$'' are broken down into concrete sources of uncertainties, the most important ones being the parametric uncertainty from the measurements of the low-energy observables $\mom_1^c$ and $\mom_2^b$, and the perturbative uncertainty reflected by the residual scale dependence. While the former is straightforward to quantify and interpret,\footnote{Note, however, a complication due to the fact that the ``experimental'' uncertainty of $\mom_n^b$ contains a contribution from perturbative QCD input for $\sqrt{s}>11.2$~GeV where no data is available at present. This leads to a large ``experimental'' uncertainty in $\mom_1^b$, and explains why the second moment $\mom_2^b$ is preferred for extracting $m_b$. The situation is expected to improve in the future.} the latter necessarily involves artificial prescriptions that may lead to bias in its estimation, as we will discuss below. We also note that the parametric uncertainty from $\alsmz$ is affected by the dependence of the extracted $m_Q(m_Q)$ on $\alsmz$, and is found to be smaller than the incorrect estimate neglecting this correlation; see Fig.~\ref{fig:PRU} below.

\begin{figure}[htb]
%\centering
\includegraphics[width=2.75in]{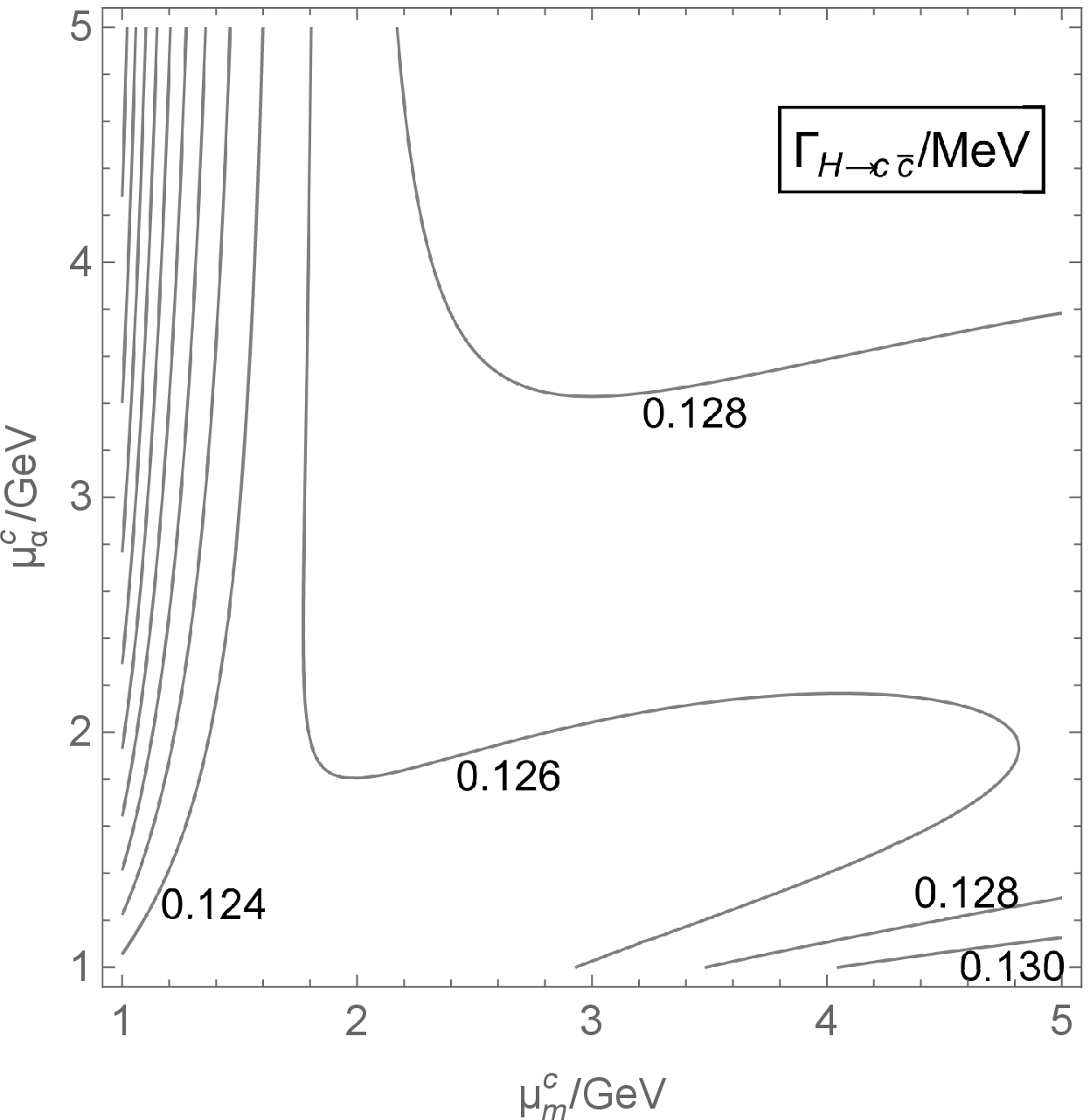}\hspace{0.25in}
\includegraphics[width=2.85in]{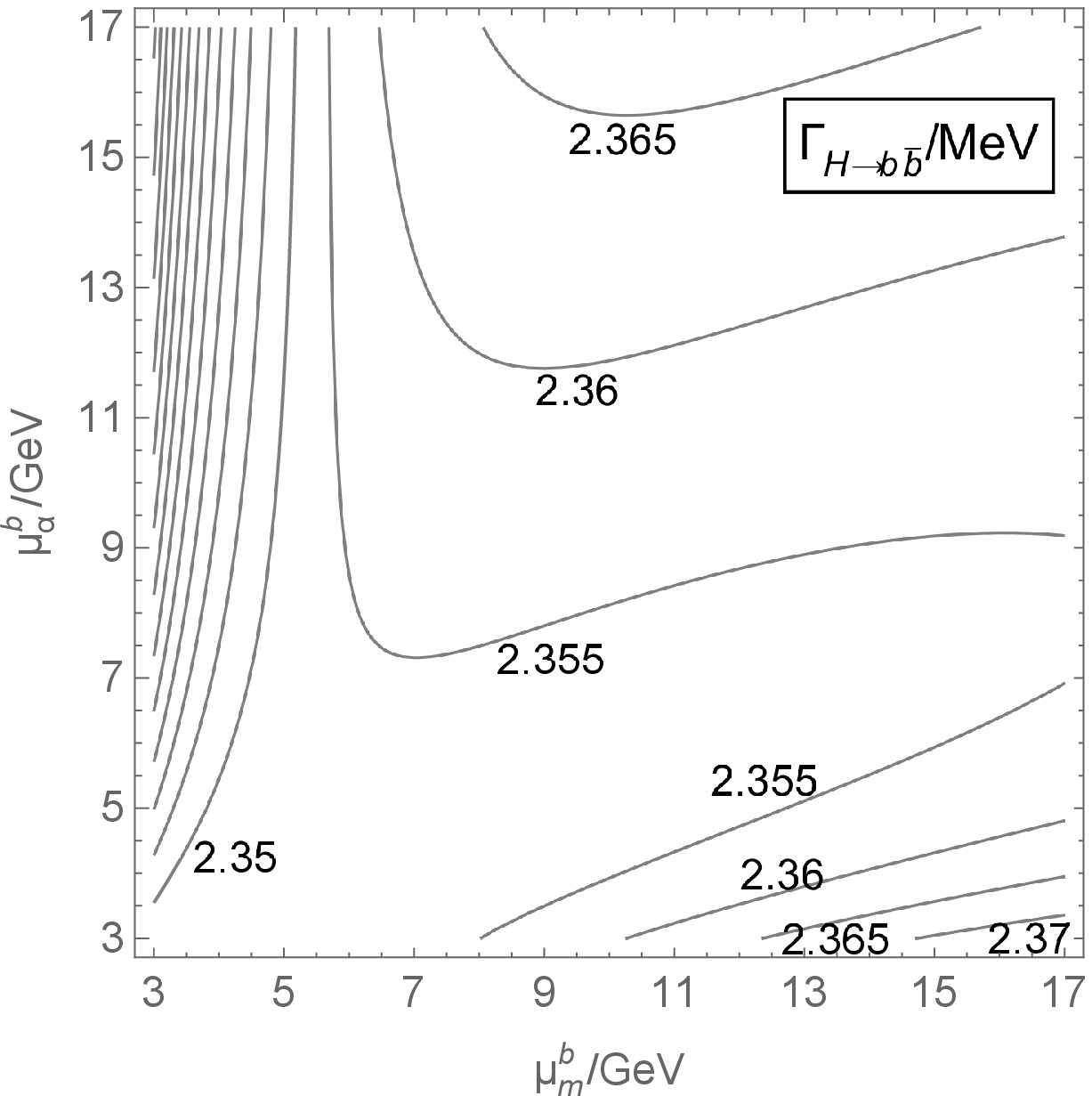}
\caption{Contours of $\Gamc$, $\Gamb$ in MeV in the $\mum$-$\mua$ plane. Unlabeled contours represent decreasing values toward the top-left corner in steps of 0.002, 0.005, respectively. The diagonal $\mum=\mua$ does not capture all the perturbative uncertainty.}
\label{fig:Gam}
\end{figure}

To visualize the perturbative uncertainty from the dependence on $\mum$, $\mua$, we fix all the input observables at their experimental central values listed in Ref.~\cite{Petrov:2015jea} and set $\mu_H=m_H$, and make contour plots for the calculated $\Gamc$, $\Gamb$ in the $\mum$-$\mua$ plane. These plots, shown in Fig.~\ref{fig:Gam}, illustrate the propagation of the scale dependence from the low-energy observables to the Higgs boson partial widths, as the latter are seen to depend on the renormalization scales chosen when calculating the former. The necessity to go beyond $\mum=\mua$ is clear since the diagonal does not capture all the scale dependence. To estimate the perturbative uncertainty, a common practice is to vary the renormalization scales within a factor of 2 around a characteristic scale of the process. But this does not directly apply to $\mum$ and $\mua$, because the $\mom_n^Q$ receive contributions from all $\sqrt{s}$; see Eq.~\leqn{momdef}. Therefore, for illustration we will vary $\mum$ and $\mua$ independently within an adjustable range $[\mumin,\mumax]$, and plot the estimated perturbative uncertainty as a function of $\mumin$ for a few choices of $\mumax$ in Fig.~\ref{fig:PRU}. The leading parametric uncertainties are shown in the same figure for comparison.

\begin{figure}[htb]
\centering
\includegraphics[width=2.75in]{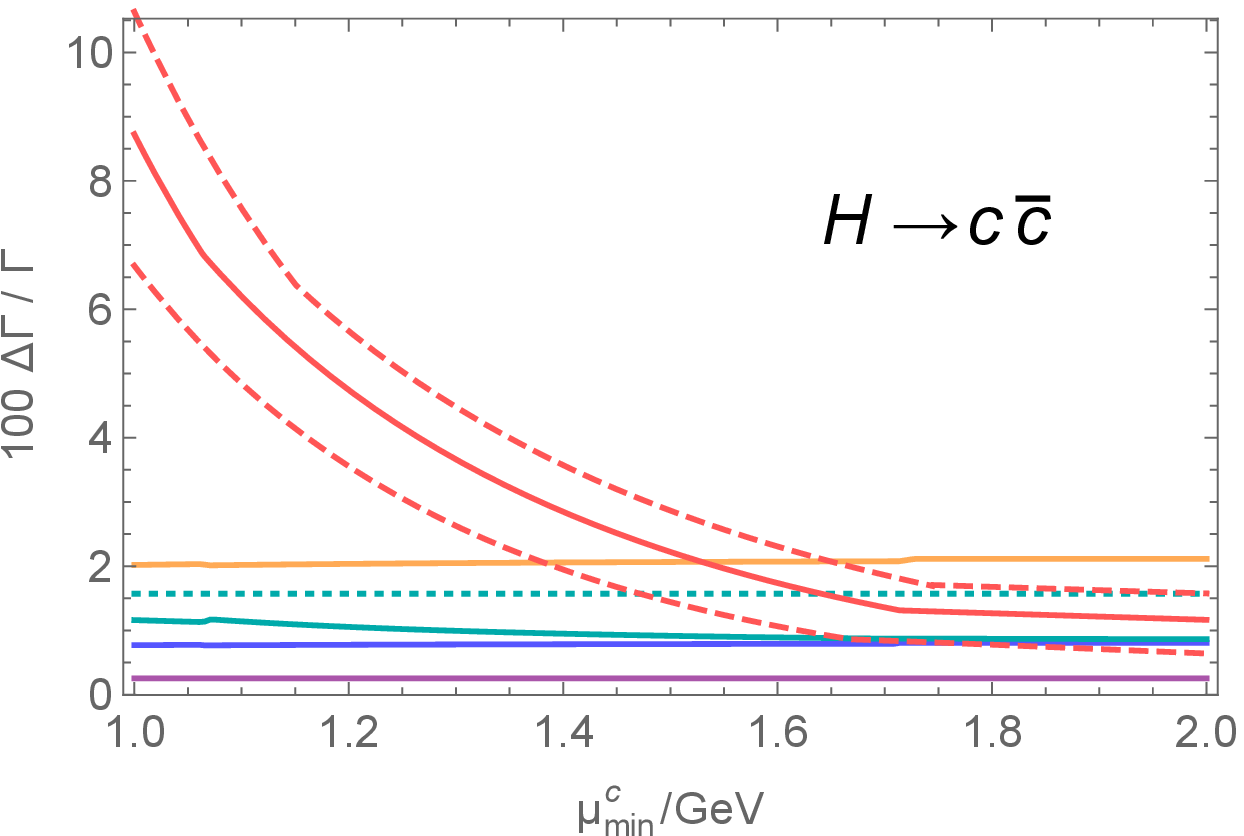}\hspace{0.25in}
\includegraphics[width=2.75in]{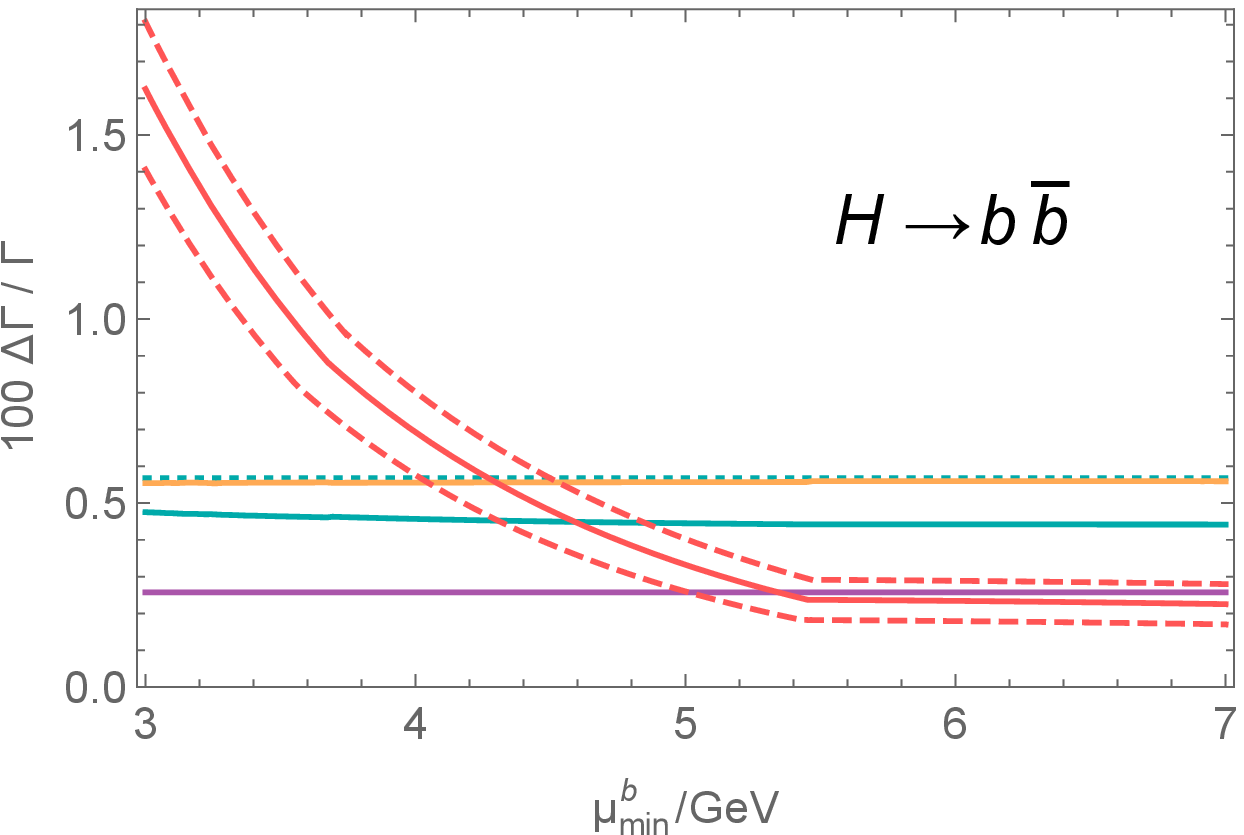}
\caption{Percent relative uncertainties in $\Gamc$, $\Gamb$ as functions of $\mumin$ from various sources: perturbative uncertainty with $\mumax^c=4$~GeV, $\mumax^b=15$~GeV (red solid) or alternatively $\mumax^c=3,5$~GeV, $\mumax^b=13,17$~GeV (red dashed), parametric uncertainties from $\mom_1^c$ or $\mom_2^b$ (orange), $\alsmz$ (cyan solid), $\mom_1^{c,\text{np}}$ (blue, for $\Gamc$ only) and $m_H$ (purple). The parametric uncertainty from $\alsmz$ incorrectly calculated assuming no correlation with $m_Q$ (cyan dotted) is also shown for comparison. See Ref.~\cite{Petrov:2015jea} for more details.}
\label{fig:PRU}
\end{figure}

It is seen from Fig.~\ref{fig:PRU} that the perturbative uncertainty from $\mum$, $\mua$ is very sensitive to the somewhat arbitrary choice of $\mumin$, and can dominate the total theory uncertainty for $\Gamc$, $\Gamb$ if lower renormalization scales are allowed in the calculations of $\mom_n^Q$. This represents a serious challenge for higher-precision Higgs boson partial widths calculations, and can be truly overcome only by calculating $\mom_n^Q$, or equivalently $\bigl(\frac{\diff}{\diff q^2}\bigr)^n\Pi_Q(q^2)\bigr|_{q^2=0}$ to higher order. Before this calculation is done, ideas on more enlightened prescriptions for the uncertainty estimation may be helpful. Two possibilities, the convergence test~\cite{Dehnadi:2015fra} and optimal scale setting~\cite{Brodsky:1982gc,Brodsky:2013vpa,Kataev:2014jba}, are discussed in Ref.~\cite{Petrov:2015jea}, but both of them are still unsatisfactory at present. It is possible that the actual situation is better in a global $\chi^2$ fit incorporating more observables in $\{\Obs{i}{low}\}$, but it remains to investigate other low-energy observables sensitive to $m_c$, $m_b$ and see if their calculations are plagued by similar scale-setting ambiguities.

In passing we briefly comment on the perturbative uncertainty from $\mu_H$, which is usually estimated by varying $\mu_H$ from $m_H/2$ to $2m_H$; see e.g.\ Ref.~\cite{Almeida:2013jfa}. It can be argued that this procedure is incomplete, since the calculations of $\Gamc$, $\Gamb$ also involve more than one renormalized parameters, and the renormalization scales chosen for $m_Q$ and $\alpha_s$ need not be equal. However, we have found that in this case, keeping the scales separate does not significantly increase the estimated uncertainty. Thus, the estimates in the literature with only one scale are expected to be robust.

\section{Conclusions}

The success of the SM is based on its agreement with data for processes across all accessible energy scales. Charm and bottom quark masses provide a bridge connecting two distinct scales $m_Q$ and $m_H$. At the next level of precision in testing the Higgs sector of the SM, theory calculations need to be improved to match the projected experimental precision. For a better understanding and a more consistent treatment of the theory uncertainties, it is desirable to resolve the information contained in $m_c$ and $m_b$ extracted from low-energy data. We have proposed a global approach involving low-energy observables and Higgs observables as a long-term goal for the precision program, and performed a first calculation that connects the two sets of observables. The analysis gives us a more detailed understanding of the ``uncertainties from $m_c$, $m_b$'', and points to future directions for the precision Higgs analysis program.

%%%%%%%%%%%%%%%%%%%%%%%%%%%%%%%%%%%%%%%%%%%%%%%%%%%%%%%%%%%%%%%%%%%%%%%%%
%%
%%   use this format to include an .eps figure into your paper
%%
%\begin{figure}[htb]
%\centering
%\includegraphics[height=3in]{figure1}
%\caption{This is how the figures should be included.}
%\label{fig:magnet}
%\end{figure}
%%%%%%%%%%%%%%%%%%%%%%%%%%%%%%%%%%%%%%%%%%%%%%%%%%%%%%%%%%%%%%%%%%%%%%%%%%%

%%%%%%%%%%%%%%%%%%%%%%%%%%%%%%%%%%%%%%%%%%%%%%%%%%%%%%%%%%%%%%%%%%%%%%%%%
%%
%%   use this format to include a LaTeX table  into your paper
%%
%\begin{table}[t]
%\begin{center}
%\begin{tabular}{l|ccc}  
%Patient &  Initial level($\mu$g/cc) &  w. Magnet &  
%w. Magnet and Sound \\ \hline
% Guglielmo B.  &   0.12     &     0.10      &     0.001  \\
% Ferrando di N. &  0.15     &     0.11      &  $< 0.0005$ \\ \hline
%\end{tabular}
%\caption{This is how the tables should be included.}
%\label{tab:blood}
%\end{center}
%\end{table}
%%%%%%%%%%%%%%%%%%%%%%%%%%%%%%%%%%%%%%%%%%%%%%%%%%%%%%%%%%%%%%%%%%%%%%%%%%%

%%%%%%%%%%%%%%%%%%%%%%%%%%%%%%%%%%
\Acknowledgements
A.A.P.\ is supported in part by the DoE under contract DE-SC0007983, Fermilab's Intensity Frontier Fellowship and URA Visiting Scholar Award \#14-S-23. 
S.P.\ is supported by the National Science Center in Poland under the research grants  DEC-2012/05/B/ST2/02597 and DEC-2012/04/A/ST2/00099.
J.D.W.\ and Z.Z.\ are supported in part by the DoE under grant DE-SC0011719.

%%%%%%%%%%%%%%%%%%%%%%%%%%%%%%%%%%


\begin{thebibliography}{99}

%%
%%  bibliographic items can be constructed using the LaTeX format in SPIRES:
%%    see    http://www.slac.stanford.edu/spires/hep/latex.html
%%  SPIRES will also supply the CITATION line information; please include it.
%%

%\cite{Gupta:2012mi}
\bibitem{Gupta:2012mi} 
  R.~S.~Gupta, H.~Rzehak and J.~D.~Wells,
  %``How well do we need to measure Higgs boson couplings?,''
  Phys.\ Rev.\ D {\bf 86}, 095001 (2012)
  [arXiv:1206.3560 [hep-ph]].
  %%CITATION = ARXIV:1206.3560;%%
  %56 citations counted in INSPIRE as of 25 août 2015

% future
%\cite{Asner:2013psa}
\bibitem{Asner:2013psa} 
  D.~M.~Asner {\it et al.},
  %``ILC Higgs White Paper,''
  arXiv:1310.0763 [hep-ph].
  %%CITATION = ARXIV:1310.0763;%%
  %126 citations counted in INSPIRE as of 25 août 2015
%\cite{Peskin:2013xra}
\bibitem{Peskin:2013xra} 
  M.~E.~Peskin,
  %``Estimation of LHC and ILC Capabilities for Precision Higgs Boson Coupling Measurements,''
  arXiv:1312.4974 [hep-ph].
  %%CITATION = ARXIV:1312.4974;%%
  %46 citations counted in INSPIRE as of 25 août 2015
%\cite{Fan:2014vta}
\bibitem{Fan:2014vta} 
  J.~Fan, M.~Reece and L.~T.~Wang,
  %``Possible Futures of Electroweak Precision: ILC, FCC-ee, and CEPC,''
  arXiv:1411.1054 [hep-ph].
  %%CITATION = ARXIV:1411.1054;%%
  %17 citations counted in INSPIRE as of 25 Aug 2015
%\cite{Ruan:2014xxa}
\bibitem{Ruan:2014xxa} 
  M.~Ruan,
  %``Higgs measurement at e+e- circular colliders,''
  arXiv:1411.5606 [hep-ex].
  %%CITATION = ARXIV:1411.5606;%%
  %8 citations counted in INSPIRE as of 25 août 2015

% theory uncertainties
%\cite{Denner:2011mq}
\bibitem{Denner:2011mq} 
  A.~Denner, S.~Heinemeyer, I.~Puljak, D.~Rebuzzi and M.~Spira,
  %``Standard Model Higgs-Boson Branching Ratios with Uncertainties,''
  Eur.\ Phys.\ J.\ C {\bf 71}, 1753 (2011)
  [arXiv:1107.5909 [hep-ph]].
  %%CITATION = ARXIV:1107.5909;%%
  %152 citations counted in INSPIRE as of 25 août 2015
%\cite{Almeida:2013jfa}
\bibitem{Almeida:2013jfa} 
  L.~G.~Almeida, S.~J.~Lee, S.~Pokorski and J.~D.~Wells,
  %``Study of the standard model Higgs boson partial widths and branching fractions,''
  Phys.\ Rev.\ D {\bf 89}, no. 3, 033006 (2014)
  [arXiv:1311.6721 [hep-ph]].
  %%CITATION = ARXIV:1311.6721;%%
  %11 citations counted in INSPIRE as of 25 août 2015
%\cite{Lepage:2014fla}
\bibitem{Lepage:2014fla} 
  G.~P.~Lepage, P.~B.~Mackenzie and M.~E.~Peskin,
  %``Expected Precision of Higgs Boson Partial Widths within the Standard Model,''
  arXiv:1404.0319 [hep-ph].
  %%CITATION = ARXIV:1404.0319;%%
  %7 citations counted in INSPIRE as of 25 Aug 2015

%\cite{Baikov:2005rw}
\bibitem{Baikov:2005rw} 
  P.~A.~Baikov, K.~G.~Chetyrkin and J.~H.~Kuhn,
  %``Scalar correlator at O(alpha(s)**4), Higgs decay into b-quarks and bounds on the light quark masses,''
  Phys.\ Rev.\ Lett.\  {\bf 96}, 012003 (2006)
  [hep-ph/0511063].
  %%CITATION = HEP-PH/0511063;%%
  %111 citations counted in INSPIRE as of 25 Aug 2015

%\cite{Agashe:2014kda}
\bibitem{Agashe:2014kda} 
  K.~A.~Olive {\it et al.} [Particle Data Group Collaboration],
  %``Review of Particle Physics,''
  Chin.\ Phys.\ C {\bf 38}, 090001 (2014).
  %%CITATION = CHPHD,C38,090001;%%
  %1731 citations counted in INSPIRE as of 25 Aug 2015

%\cite{Petrov:2015jea}
\bibitem{Petrov:2015jea} 
  A.~A.~Petrov, S.~Pokorski, J.~D.~Wells and Z.~Zhang,
  %``Role of low-energy observables in precision Higgs boson analyses,''
  Phys.\ Rev.\ D {\bf 91}, no. 7, 073001 (2015)
  [arXiv:1501.02803 [hep-ph]].
  %%CITATION = ARXIV:1501.02803;%%
  %3 citations counted in INSPIRE as of 25 août 2015

% low moments
%\cite{Dehnadi:2011gc}
\bibitem{Dehnadi:2011gc} 
  B.~Dehnadi, A.~H.~Hoang, V.~Mateu and S.~M.~Zebarjad,
  %``Charm Mass Determination from QCD Charmonium Sum Rules at Order $\alpha_{s}^{3}$,''
  JHEP {\bf 1309}, 103 (2013)
  [arXiv:1102.2264 [hep-ph]].
  %%CITATION = ARXIV:1102.2264;%%
  %47 citations counted in INSPIRE as of 26 Aug 2015
%\cite{Kuhn:2001dm}
\bibitem{Kuhn:2001dm} 
  J.~H.~Kuhn and M.~Steinhauser,
  %``Determination of alpha(s) and heavy quark masses from recent measurements of R(s),''
  Nucl.\ Phys.\ B {\bf 619}, 588 (2001)
  [Nucl.\ Phys.\ B {\bf 640}, 415 (2002)]
  [hep-ph/0109084].
  %%CITATION = HEP-PH/0109084;%%
  %144 citations counted in INSPIRE as of 26 août 2015
%\cite{Kuhn:2007vp}
\bibitem{Kuhn:2007vp} 
  J.~H.~Kuhn, M.~Steinhauser and C.~Sturm,
  %``Heavy Quark Masses from Sum Rules in Four-Loop Approximation,''
  Nucl.\ Phys.\ B {\bf 778}, 192 (2007)
  [hep-ph/0702103 [HEP-PH]].
  %%CITATION = HEP-PH/0702103;%%
  %144 citations counted in INSPIRE as of 26 août 2015
%\cite{Chetyrkin:2009fv}
\bibitem{Chetyrkin:2009fv} 
  K.~G.~Chetyrkin, J.~H.~Kuhn, A.~Maier, P.~Maierhofer, P.~Marquard, M.~Steinhauser and C.~Sturm,
  %``Charm and Bottom Quark Masses: An Update,''
  Phys.\ Rev.\ D {\bf 80}, 074010 (2009)
  [arXiv:0907.2110 [hep-ph]].
  %%CITATION = ARXIV:0907.2110;%%
  %175 citations counted in INSPIRE as of 26 août 2015
%\cite{Dehnadi:2015fra}
\bibitem{Dehnadi:2015fra} 
  B.~Dehnadi, A.~H.~Hoang and V.~Mateu,
  %``Bottom and Charm Mass Determinations with a Convergence Test,''
  arXiv:1504.07638 [hep-ph].
  %%CITATION = ARXIV:1504.07638;%%
  %1 citations counted in INSPIRE as of 26 août 2015

% high moments
%\cite{Signer:2008da}
\bibitem{Signer:2008da} 
  A.~Signer,
  %``The Charm quark mass from non-relativistic sum rules,''
  Phys.\ Lett.\ B {\bf 672}, 333 (2009)
  [arXiv:0810.1152 [hep-ph]].
  %%CITATION = ARXIV:0810.1152;%%
  %18 citations counted in INSPIRE as of 26 août 2015
%\cite{Hoang:2012us}
\bibitem{Hoang:2012us} 
  A.~Hoang, P.~Ruiz-Femenia and M.~Stahlhofen,
  %``Renormalization Group Improved Bottom Mass from Upsilon Sum Rules at NNLL Order,''
  JHEP {\bf 1210}, 188 (2012)
  [arXiv:1209.0450 [hep-ph]].
  %%CITATION = ARXIV:1209.0450;%%
  %25 citations counted in INSPIRE as of 26 août 2015
%\cite{Penin:2014zaa}
\bibitem{Penin:2014zaa} 
  A.~A.~Penin and N.~Zerf,
  %``Bottom Quark Mass from $\Upsilon$ Sum Rules to ${\cal O}(\alpha_s^3)$,''
  JHEP {\bf 1404}, 120 (2014)
  [arXiv:1401.7035 [hep-ph]].
  %%CITATION = ARXIV:1401.7035;%%
  %16 citations counted in INSPIRE as of 26 août 2015
%\cite{Beneke:2014pta}
\bibitem{Beneke:2014pta} 
  M.~Beneke, A.~Maier, J.~Piclum and T.~Rauh,
  %``The bottom-quark mass from non-relativistic sum rules at NNNLO,''
  Nucl.\ Phys.\ B {\bf 891}, 42 (2015)
  [arXiv:1411.3132 [hep-ph]].
  %%CITATION = ARXIV:1411.3132;%%
  %9 citations counted in INSPIRE as of 26 août 2015

% variants of moments
%\cite{Bodenstein:2011ma}
\bibitem{Bodenstein:2011ma} 
  S.~Bodenstein, J.~Bordes, C.~A.~Dominguez, J.~Penarrocha and K.~Schilcher,
  %``QCD sum rule determination of the charm-quark mass,''
  Phys.\ Rev.\ D {\bf 83}, 074014 (2011)
  [arXiv:1102.3835 [hep-ph]].
  %%CITATION = ARXIV:1102.3835;%%
  %20 citations counted in INSPIRE as of 26 août 2015
%\cite{Bodenstein:2011fv}
\bibitem{Bodenstein:2011fv} 
  S.~Bodenstein, J.~Bordes, C.~A.~Dominguez, J.~Penarrocha and K.~Schilcher,
  %``Bottom-quark mass from finite energy QCD sum rules,''
  Phys.\ Rev.\ D {\bf 85}, 034003 (2012)
  [arXiv:1111.5742 [hep-ph]].
  %%CITATION = ARXIV:1111.5742;%%
  %22 citations counted in INSPIRE as of 26 Aug 2015

% semileptonic B decay
%\cite{Gambino:2013rza}
\bibitem{Gambino:2013rza} 
  P.~Gambino and C.~Schwanda,
  %``Inclusive semileptonic fits, heavy quark masses, and $V_{cb}$,''
  Phys.\ Rev.\ D {\bf 89}, no. 1, 014022 (2014)
  [arXiv:1307.4551 [hep-ph]].
  %%CITATION = ARXIV:1307.4551;%%
  %42 citations counted in INSPIRE as of 26 Aug 2015
%\cite{Buchmuller:2005zv}
\bibitem{Buchmuller:2005zv} 
  O.~Buchmuller and H.~Flacher,
  %``Fit to moment from B ---> X(c) l anti-nu and B ---> X(s) gamma decays using heavy quark expansions in the kinetic scheme,''
  Phys.\ Rev.\ D {\bf 73}, 073008 (2006)
  [hep-ph/0507253].
  %%CITATION = HEP-PH/0507253;%%
  %192 citations counted in INSPIRE as of 26 août 2015
%\cite{Bauer:2004ve}
\bibitem{Bauer:2004ve} 
  C.~W.~Bauer, Z.~Ligeti, M.~Luke, A.~V.~Manohar and M.~Trott,
  %``Global analysis of inclusive B decays,''
  Phys.\ Rev.\ D {\bf 70}, 094017 (2004)
  [hep-ph/0408002].
  %%CITATION = HEP-PH/0408002;%%
  %174 citations counted in INSPIRE as of 26 Aug 2015

% sum rules
%\cite{Novikov:1977dq}
\bibitem{Novikov:1977dq} 
  V.~A.~Novikov, L.~B.~Okun, M.~A.~Shifman, A.~I.~Vainshtein, M.~B.~Voloshin and V.~I.~Zakharov,
  %``Charmonium and Gluons: Basic Experimental Facts and Theoretical Introduction,''
  Phys.\ Rept.\  {\bf 41}, 1 (1978).
  %%CITATION = PRPLC,41,1;%%
  %762 citations counted in INSPIRE as of 26 Aug 2015
%\cite{Shifman:1998rb}
\bibitem{Shifman:1998rb} 
  M.~A.~Shifman,
  %``Snapshots of hadrons or the story of how the vacuum medium determines the properties of the classical mesons which are produced, live and die in the QCD vacuum,''
  Prog.\ Theor.\ Phys.\ Suppl.\  {\bf 131}, 1 (1998)
  [hep-ph/9802214].
  %%CITATION = HEP-PH/9802214;%%
  %92 citations counted in INSPIRE as of 26 Aug 2015
%\cite{Colangelo:2000dp}
\bibitem{Colangelo:2000dp} 
  P.~Colangelo and A.~Khodjamirian,
  %``QCD sum rules, a modern perspective,''
  In *Shifman, M. (ed.): At the frontier of particle physics, vol. 3* 1495-1576
  [hep-ph/0010175].
  %%CITATION = HEP-PH/0010175;%%
  %422 citations counted in INSPIRE as of 26 Aug 2015

% lattice
%\cite{McNeile:2010ji}
\bibitem{McNeile:2010ji} 
  C.~McNeile, C.~T.~H.~Davies, E.~Follana, K.~Hornbostel and G.~P.~Lepage,
  %``High-Precision c and b Masses, and QCD Coupling from Current-Current Correlators in Lattice and Continuum QCD,''
  Phys.\ Rev.\ D {\bf 82}, 034512 (2010)
  [arXiv:1004.4285 [hep-lat]].
  %%CITATION = ARXIV:1004.4285;%%
  %168 citations counted in INSPIRE as of 26 Aug 2015
%\cite{Chakraborty:2014aca}
\bibitem{Chakraborty:2014aca} 
  B.~Chakraborty {\it et al.},
  %``High-precision quark masses and QCD coupling from $n_f=4$ lattice QCD,''
  Phys.\ Rev.\ D {\bf 91}, no. 5, 054508 (2015)
  [arXiv:1408.4169 [hep-lat]].
  %%CITATION = ARXIV:1408.4169;%%
  %13 citations counted in INSPIRE as of 26 août 2015
%\cite{Colquhoun:2014ica}
\bibitem{Colquhoun:2014ica} 
  B.~Colquhoun, R.~J.~Dowdall, C.~T.~H.~Davies, K.~Hornbostel and G.~P.~Lepage,
  %``$\Upsilon$ and $\Upsilon^{\prime}$ Leptonic Widths, $a_{\mu}^b$ and $m_b$ from full lattice QCD,''
  Phys.\ Rev.\ D {\bf 91}, no. 7, 074514 (2015)
  [arXiv:1408.5768 [hep-lat]].
  %%CITATION = ARXIV:1408.5768;%%
  %13 citations counted in INSPIRE as of 26 Aug 2015

% \mom_n^Q calculations
%\cite{Maier:2009fz}
\bibitem{Maier:2009fz} 
  A.~Maier, P.~Maierhofer, P.~Marquard and A.~V.~Smirnov,
  %``Low energy moments of heavy quark current correlators at four loops,''
  Nucl.\ Phys.\ B {\bf 824}, 1 (2010)
  [arXiv:0907.2117 [hep-ph]].
  %%CITATION = ARXIV:0907.2117;%%
  %45 citations counted in INSPIRE as of 26 août 2015
%\cite{Broadhurst:1994qj}
\bibitem{Broadhurst:1994qj} 
  D.~J.~Broadhurst, P.~A.~Baikov, V.~A.~Ilyin, J.~Fleischer, O.~V.~Tarasov and V.~A.~Smirnov,
  %``Two loop gluon condensate contributions to heavy quark current correlators: Exact results and approximations,''
  Phys.\ Lett.\ B {\bf 329}, 103 (1994)
  [hep-ph/9403274].
  %%CITATION = HEP-PH/9403274;%%
  %84 citations counted in INSPIRE as of 26 Aug 2015

%\cite{Chetyrkin:2000yt}
\bibitem{Chetyrkin:2000yt} 
  K.~G.~Chetyrkin, J.~H.~Kuhn and M.~Steinhauser,
  %``RunDec: A Mathematica package for running and decoupling of the strong coupling and quark masses,''
  Comput.\ Phys.\ Commun.\  {\bf 133}, 43 (2000)
  [hep-ph/0004189].
  %%CITATION = HEP-PH/0004189;%%
  %229 citations counted in INSPIRE as of 09 sept. 2015

% BLM
%\cite{Brodsky:1982gc}
\bibitem{Brodsky:1982gc} 
  S.~J.~Brodsky, G.~P.~Lepage and P.~B.~Mackenzie,
  %``On the Elimination of Scale Ambiguities in Perturbative Quantum Chromodynamics,''
  Phys.\ Rev.\ D {\bf 28}, 228 (1983).
  %%CITATION = PHRVA,D28,228;%%
  %1018 citations counted in INSPIRE as of 27 août 2015
%\cite{Brodsky:2013vpa}
\bibitem{Brodsky:2013vpa} 
  S.~J.~Brodsky, M.~Mojaza and X.~G.~Wu,
  %``Systematic Scale-Setting to All Orders: The Principle of Maximum Conformality and Commensurate Scale Relations,''
  Phys.\ Rev.\ D {\bf 89}, no. 1, 014027 (2014)
  [arXiv:1304.4631 [hep-ph]].
  %%CITATION = ARXIV:1304.4631;%%
  %41 citations counted in INSPIRE as of 27 Aug 2015
%\cite{Kataev:2014jba}
\bibitem{Kataev:2014jba} 
  A.~L.~Kataev and S.~V.~Mikhailov,
  %``Generalization of the Brodsky-Lepage-Mackenzie optimization within the {?}-expansion and the principle of maximal conformality,''
  Phys.\ Rev.\ D {\bf 91}, no. 1, 014007 (2015)
  [arXiv:1408.0122 [hep-ph]].
  %%CITATION = ARXIV:1408.0122;%%
  %6 citations counted in INSPIRE as of 27 Aug 2015

\end{thebibliography}
\end{document}